\documentclass[11pt]{article}
\usepackage{array}
\usepackage{graphicx}
\usepackage{cite}
\usepackage{textpos}

\title{Superfluidity and Space-Time Translation Symmetry Breaking}

\author{Frank Wilczek\\
\small\it Center for Theoretical Physics, MIT, Cambridge MA 02139 USA}

\begin{document}

\maketitle

\begin{textblock*}{5cm}(11cm,-8.2cm)
 \fbox{\footnotesize MIT-CTP-4486}
 \end{textblock*}

\begin{abstract}
I present a simple model that exhibits a temporal analogue of superconducting crystalline (LOFF) ordering.  I sketch designs for minimally dissipative AC circuits, all based on time translation symmetry ($\tau$) invariant dynamics, exploiting weak links (Josephson effects).   These systems violate $\tau$ spontaneously.  I also discuss effective theories of that phenomenon, and space-time generalizations. 
\end{abstract}

\medskip

Recently there has been considerable interest in the possibility of spontaneous breaking of time translation symmetry $\tau$ \cite{quantumTXtals}-\cite{reply}.   Here I bring in ideas from superfluidity which offer additional perspectives and widen the possibilities significantly.   

{\it Microscopic Model}:  The energy functional
\begin{equation}\label{loffE}
E_{\rm loff} ~\propto~ \frac{1}{4} | \vec \partial \phi - \frac{2e}{\hbar} \vec A |^4 - \frac{\zeta}{2} | \vec \partial \phi - \frac{2e}{\hbar} \vec A |^2
\end{equation}
describes superconductors with wave-like or crystalline (Larkin-Ovchinnikov-Ferrell-Fulde \cite{LO} \cite{FF}, or ``LOFF'') condensates.  In the absence of a vector potential we will have, upon minimizing the energy in Eqn.\,(\ref{loffE}),
\begin{equation}\label{spaceVariation}
\phi (\vec x) ~=~   \vec k \cdot \vec x + \phi_0
\end{equation} 
for some wavevector $\vec k$ with $k^2 = \zeta$.  Several possible realizations of LOFF states have been identified \cite{matsuda}-\cite{bianchi}.

It is natural to consider the possibility of a temporal analogue to the spatial behavior indicated in Eqn.\,(\ref{spaceVariation}).  
What we want, is that the pairing occurs between states separated by a characteristic frequency.   For superconducting systems the absolute frequency dependence is rendered ambiguous by the possibility of time-dependent gauge transformations, or stated more simply by the lack of a natural zero of energy, so it is simplest to use the language of particle-hole pairing.  

For orientation purposes, let us begin by further specializing to the transparent case of two flat bands with energies $\varepsilon_1 <  \varepsilon_2$, and the Hamiltonian
\begin{eqnarray}\label{modelH}
H ~&=&~ \frac{\varepsilon_2 + \varepsilon_1}{2} \, N \, + \, (\varepsilon_2 - \varepsilon_1)\, S_3 \, - \, g \, (S_-  S_+ \, + \, S_+ S_-)  \nonumber \\
~&=&~ \frac{\varepsilon_2 + \varepsilon_1}{2} \, N \, + \, (\varepsilon_2 - \varepsilon_1)\, S_3 \, - \, 2g \, (\vec S^2 - S_3^2)
\end{eqnarray}
where 
\begin{equation}
N ~=~  \sum\limits_k b_k^\dagger b_k + \sum\limits_k a_k^\dagger a_k
\end{equation}
is the total occupation number and 
\begin{eqnarray}
S_+ ~&=&~ \sum\limits_k b_k^\dagger a_k \nonumber \\
~&=&~ S_1 + i S_2 \nonumber \\
~&=&~ S_-^\dagger \nonumber \\
S_3 ~&=&~ \frac{1}{2} ( \sum\limits_k b_k^\dagger b_k - \sum\limits_k a_k^\dagger a_k ) 
\end{eqnarray} 
define hermitean pseudospin operators $S_1, S_2, S_3$ that satisfy the algebra of angular momentum and generate isospin-like rotations between the $a$ and $b$ modes.

Since $N, \vec S^2,$ and $S_3$ commute, we can construct the minimum energy states for $H$, given $N$, by maximizing $S$ (so $S = N/2$) and choosing a state with definite $S_3$ 
If $S_3$ is also allowed to vary, the minimum will occur for
\begin{equation}\label{s3Condition}
\langle S_3 \rangle ~=~ {\rm Max} \, (- \frac{\varepsilon_2 - \varepsilon_1}{4g} , - \frac{N}{2} )
\end{equation}
(The second alternative, which saturates the population of the $a$ modes, is essentially trivial.)
On the other hand it can be appropriate to hold the expectation value of $S_3$ fixed.  We can imagine, for example, that the $a$ and $b$ modes correspond to states in distinct layers, whose total occupations can be fixed independently.  (Note that the assumed interaction term does not require interlayer tunneling -- this is just another way of saying that it commutes with $S_3$.)  As we can see by re-arranging 
\begin{equation}
a^\dagger_k b_k b^\dagger_l a_l ~\stackrel{\sim}{\rightarrow}~ - a^\dagger_k a_l b^\dagger_l b_k
\end{equation}
the assumed interaction corresponds, roughly, to an effective repulsion between density waves that does not depend on momentum transfer.

Now we can follow the classic BCS procedure, postulating a symmetry-breaking condensate.  In this procedure, we assume the {\it ansatz\/} 
\begin{equation}\label{condensate}
\langle \mu, \theta | S_+ | \mu, \theta \rangle ~=~ \Delta_0 e^{i\theta}
\end{equation}
with $\Delta_0$ a number, ultimately fixed self-consistently by the gap equation.
Since 
\begin{equation}
\lbrack H, S_+ \rbrack ~=~  (\varepsilon_2 - \varepsilon_1)\, S_+ \, + \, 2g \, (S_3 S_+ \, + \, S_+ S_3 )
\end{equation}
and
\begin{equation}
\frac{d}{dt} \langle  S_+  \rangle ~=~ i \langle  \lbrack H, S_+ \rbrack   \rangle 
\end{equation}
consistent classical evolution for $\theta$ requires
\begin{equation}
\dot \theta ~=~  \varepsilon_2 - \varepsilon_1 + 4g \langle S_3 \rangle
\end{equation}
This vanishes if $ \langle S_3 \rangle$ is fixed non-trivially by Eqn.\,(\ref{s3Condition}), but not if that expectation value is pinned at a different value.

In the pseudospin formalism, this time dependence has a simple interpretation:  The condensate is an effective spin of fixed magnitude at a fixed angle to the $\hat z$ axis,  and the $\varepsilon_2 - \varepsilon_1$ term supplies an effective magnetic field in the $\hat z$ direction, which induces precession.  

The BCS condensation {\it ansatz\/} is overkill for this simple model (where all  states with the total spin and expectation value of $S_3$ are degenerate eigenstates).  Its virtue is its ability, at the price of more complicated algebra, to accommodate more complex, momentum-dependent energies and interactions than assumed in Eqn.\,(\ref{modelH}).   One expects  qualitative aspects of spontaneous symmetry breaking to survive such generalizations.    One can also consider bosonic systems along the same lines.  Indeed, related techniques have been applied to discuss dynamic magnon condensation in liquid He$^3$ \cite{volovik} and density oscillations in two-component cold atomic gases \cite{wiemann} \cite{kasamatsu}.  In both those contexts, very long-lived oscillatory states have been observed.  

In principle one can probe for LOFF behavior, or its temporal analogue, by comparing the phase relationships among currents induced by weak-link contacts with a conventional superfluid at several points, an idea elaborated below and in \cite{Xiong}.

{\it $\tau$ Breaking - Separation of Dynamics and Measurement}: The AC Josephson effect gives an example of time dependent, and thus time-translation symmetry breaking, behavior in a system specified by time-independent conditions.  Since its standard realization involves continual (alternating) current flow across a voltage, however, it describes behavior in a dissipative system.   Here I will emphasize, and then build upon, the simple observation that by making the connection intermittent -- thus regarding it as a probe, rather than an intrinsic part, of the dynamics -- we can make the dissipation arbitrarily small, while retaining the time dependence.   This supplies us both an example of a well understood system that spontaneously breaks time-translation symmetry (``time crystal'') and, when generalized, a technique for probing possible novel dynamical realizations.    I will also describe another, perhaps more elegant, way to avoid dissipation, using two weak links.  

Since they are basic to everything that follows, a quick recollection of the Josephson phenomena is in order.  (For a simple conceptual introduction see \cite{Feynman}; for a more sophisticated introduction to the state of the art see \cite{Tinkham}.)   We consider two bulk superconductors connected by a weak link; for simplicity we suppose that the contact is localized, so that the spatial variation of the phases $\theta_1, \theta_2$ of the two superconductors near the contact, and the vector potential across the link, can be neglected.   Then the basic Josephson relations are
\begin{eqnarray}
\frac{d \delta}{dt} ~&=&~ \frac{2eV}{\hbar} \label{ACJosephson} \\
j  ~&=&~ \eta g(\delta) \label{DCJosephson}
\end{eqnarray}
where $\delta \equiv \theta_2 - \theta_1 $ is the relative phase, $V$ is the voltage across the junction, $g(\delta)$ is a non-trivial $2\pi$-periodic function often approximated as $\sin \delta$, and $\eta$ is a coupling parameter introduced for later convenience.   

Now if $V$ and $\eta$ are non-zero constants then according to Eqns.\,({\ref{ACJosephson}, \ref{DCJosephson}) we will have the time dependent current 
\begin{equation}\label{tCurrent}
j(t) ~=~ \eta \, g(\frac{2eV}{\hbar}t + \delta_0)
\end{equation}
where $\delta_0$ is an integration constant.  This presents a manifestly time dependent physical phenomenon, though nothing in the specification of the problem broke time translation symmetry.  In that sense it is an example of spontaneous $\tau$ breaking.   The occurrence of an undetermined parameter (``soft mode'') $\delta_0$ within a manifold of solutions fits that interpretation.   

On the other hand the movement of charge, in the presence of a potential difference, will involve dissipation, and in a closed system with normal (nonsuperconducting) elements to close the circuit.  If there is no external energy source of energy to sustain it,  $V$ will relax to zero.   So if our ambition is to exhibit highly persistent ``ground state'' spontaneous time-translation symmetry, the standard AC Josephson effect does not quite serve.   

That objection, however, is more formal than substantial.  We can cleanly separate the conceptually time dependent effect, Eqn.\,(\ref{ACJosephson}), from its practical manifestation Eqn.\,(\ref{DCJosephson}).   Specifically, by making and breaking contact we can arrange $\eta \rightarrow \eta(t)$ to vanish except at designated ``measurement'' times, and to be small even then.  In other words, we can choose to regard the separated superconductors as the system of interest, and the junction as a measuring device. Then in the ground state we will have the time dependent relation Eqn.\,(\ref{tCurrent}), which entails measurable physical consequences (and contains $\eta (t)$ only as a multiplicative factor), in a system with arbitrarily small dissipation.  Practical implementation of a low-dissipation switch in this context raises several challenging issues, as discussed in the companion paper \cite{Xiong}, where a concrete design is proposed.  (The design employs extended, as opposed to point, contacts, so more complicated equations apply.)

One can avoid normal components altogether, by closing the circuit in an annular arrangement with a second weak link.   Ideally, if $g(\delta) = \sin \delta$, the junctions are identical, and a magnetic flux of magnitude 
$h/4e$ threads the annulus, then the AC Josephson currents at the junctions will be equal and opposite, due to a phase offset $\pi$, thus closing the circuit.  If those idealized conditions are met approximately, then plausibly the system will settle into a mode of operation wherein some charge accumulates (DC) at one or both junctions and an appropriate steady (DC) edge supercurrent supplies corrective flux, so that there is no net charge accumulation per cycle.  In this mode, we realize dissipation-free yet time dependent current flow.  (I am neglecting electromagnetic radiation, whose effect can be minimized in several ways \cite{quantumTXtals}).   Related frequency locking phenomena in Josephson arrays have been discussed previously \cite{wcs}.  

Since the voltage difference can in principle relax, unless special precautions are taken, one may choose not to regard this as a strictly ground-state phenomenon, in which case we could speak of a minimally dissipative $\tau$ symmetry breaking system, as distinct from a strict time crystal.   (To me, whether that is a useful distinction seems somewhat a matter of taste, hinging on to what extent one is willing to regard the special precautions as intrinsic to defining ``the system'').  

{\it Effective Theory}: Let us now adopt a broader perspective, to consider the possible implications of less conventional dynamics for superconductor 2.   The effect of this will be to modify Eqn.(\ref{ACJosephson}).    To set the stage for generalizations, let us recall the default assumptions, which lead to Eqn.\,(\ref{ACJosephson}), in a way suggestive for our purposes.  The energy functional of the superconductors contain terms of the form
\begin{equation}
E_{\rm minimal} ~\propto~ (\dot \theta - \frac{2e}{\hbar} A_0)^2  
\end{equation}
This form is consistent with the appropriate gauge symmetry
\begin{eqnarray}
\theta^\prime ~&=&~ \theta + \frac{2e}{\hbar} \lambda (t) \nonumber \\
A_0^\prime ~&=&~ A_0 + \dot \lambda (t)
\end{eqnarray}
If each superconductor minimizes an energy functional of this type, then Eqn.\,(\ref{ACJosephson}) follows.  

On the other hand, suppose that the energy functional of superconductor 2 is of a less conventional type, suggested by an extension of the Landau-Ginzburg philosophy, in the form
\begin{equation}\label{motiveE}
E_{\rm motive} ~\propto~ \frac{3}{4} (\dot \theta - \frac{2e}{\hbar} A_0)^4 - \frac{\kappa}{2} (\dot \theta - \frac{2e}{\hbar} A_0)^2
\end{equation}
with $\kappa > 0$, while superconductor 1 is conventional.  (The factor 3 is adopted for consistency with \cite{classicalTXtal}.) Then we will have, in place of Eqn.\,(\ref{ACJosephson}),
\begin{equation}\label{ACTXtal}
\frac{d \delta}{dt} ~=~ \frac{2eV}{\hbar} \, \pm \, \sqrt{\frac{\kappa}{3}}
\end{equation}
In principle, this behavior might be probed by use of Eqn.\,(\ref{DCJosephson}), with a small intermittent $\eta$.   Note that Eqn.\,(\ref{ACTXtal}) remains non-trivial for $V=0$, giving a dissipationless time crystal.  Unfortunately practical identification is complicated by the possibility of non-trivial internal potentials, which can also contribute to $V$.  The bifurcation of frequencies could be a more robust characteristic.

The condensate in a superconductor supporting both kinds of unconventional terms, Eqn.\,(\ref{motiveE}, \ref{loffE}), would exhibit traveling waves in $\phi$, realizing a space-time crystal.  A LOFF superconductor subject to a non-zero potential $V$ would also serve for that purpose.   



{\it Spatial Josephson Effect -- Significance of Vector Potential Offset}: It is interesting, and falls naturally within our exploration of time$\leftrightarrow$space analogies, to consider the possibility of a spatial analog of Eqn.\,(\ref{ACJosephson}), in the form
\begin{equation}\label{spatialJosephson}
\frac{d\delta}{dz} ~=~ \frac{2e}{\hbar} \, (A_z (2) - A_z (1) ) 
\end{equation}
Just as jumps in $A_0$ can be imprinted by parallel capacitor plates with opposite charge densities, jumps in $A_z$ can be imprinted by parallel current sheets with opposite $j_z$.  If we imagine two superconductors on opposite sides of the $x = 0$ plane, where such parallel current sheets are found, then Eqn.\,(\ref{spatialJosephson}) will apply.  (Of course, one will have to allow for small windows in the current sheets, where weak links can form.)  

This effect exhibits a direct physical significance for vector potential offsets, similar in spirit to the Aharonov-B\"ohm effect.  Indeed, if we draw a loop with short lines connecting the two superconductors at $z = z_a, z_b$ near $x=0$, and joined up by lines inside the superconductors, the enclosed magnetic flux will be $(A_z(2) - A_z(1) ) \cdot (z_b - z_a)$, and the change in $\delta$ specified by Eqn.\,(\ref{spatialJosephson}) reflects that flux directly.   (Integral of $F_{xz}$ in the $xz$ plane.) The conventional AC Josephson effect can be interpreted in a similar way, but now involving loops in the $xt$ plane and enclosed electric flux. (Integral of $F_{xt}$ in the $xt$ plane.)  One can, of course, combine the effects.

{\it Conclusions}: Temporal, and mixed spatio-temporal analogues of LOFF can arise in microscopic models that plausibly might correspond to realizable systems.  It appears to be possible to exhibit time-dependent phenomena in time independent systems that are asymptotically free of dissipation, in the limit of infrequent measurement, or very nearly so. The preceding discussion of weak links is a concrete embodiment of the framing of the issue of observability of time-translation symmetry in \cite{quantumTXtals} and the related discussions in \cite{TXtalsExpt}.  I have  emphasized the language of superconductivity, but the central idea, that weak links can be used to probe unconventional order parameter dynamics, is more general.

\bigskip

{\it Acknowledgement}: I am grateful to C.~Nayak, A.~Shapere, Z.~Xiong, and M.~Hertzberg for helpful discussions. 
This work is supported by the U.S. Department of Energy under contract No. DE-FG02-05ER41360.

\end{document}